\documentclass{article}
\usepackage{jheppub}

\let\ifpdf\relax
\usepackage{ifpdf}
\usepackage{amsmath,amssymb,graphicx}
\usepackage{longtable}
\usepackage{verbatim}
\usepackage{float}

\usepackage{color}

\usepackage[usenames,dvipsnames]{xcolor}
\definecolor{darkerblue}{rgb}{0.2,0.2,0.5}

\usepackage{tikz}
\usetikzlibrary{trees}
\usetikzlibrary{decorations.pathmorphing}
\usetikzlibrary{decorations.markings}
\tikzset{
    photon/.style={decorate, decoration={snake}, draw=black},
    wino/.style={draw=redwine},    
    electron/.style={draw=black, postaction={decorate},
        decoration={markings,mark=at position .55 with {\arrow[draw=black]{>}}}},
    scalar/.style={draw=black, dashed,postaction={decorate},
        decoration={markings,mark=at position .55 with {\arrow[draw=black]{>}}}},
    gluon/.style={decorate, draw=black,
        decoration={coil,amplitude=4pt, segment length=5pt}}
}

\usepackage{afterpage}

\newcommand{\met}{E\!\!\!/_T}

\newcommand{\bear}{\begin{array}}
\newcommand{\ear}{\end{array}}

\newcommand{\beq}{\begin{eqnarray}}
\newcommand{\eeq}{\end{eqnarray}}
\newcommand{\beqa}{\begin{eqnarray}}
\newcommand{\eeqa}{\end{eqnarray}}

\def\OMIT#1{{}}
\newcommand{\lsim}{\mathrel{\rlap{\lower4pt\hbox{\hskip1pt$\sim$}}
    \raise1pt\hbox{$<$}}}         
\newcommand{\gsim}{\mathrel{\rlap{\lower4pt\hbox{\hskip1pt$\sim$}}
    \raise1pt\hbox{$>$}}}         

\begin{document}

\title{Pseudoscalar Portal Dark Matter and New Signatures of Vector-like Fermions}

\author{JiJi Fan,} 
\author{Savvas M. Koushiappas,}
\author{and Greg Landsberg}
\affiliation{Department of Physics, Brown University, Providence, RI, 02912, USA}

\abstract{Fermionic dark matter interacting with the Standard Model sector through a pseudoscalar portal could evade the direct detection constraints while preserving a WIMP miracle. We study the LHC constraints on the pseudoscalar production in simplified models with the pseudoscalar either dominantly coupled to $b$ quarks or $\tau$ leptons and explore their implications for the GeV excesses in gamma ray observations. We also investigate models with new vector-like fermions that could realize the simplified models of pseudoscalar portal dark matter. These models yield new decay channels and signatures of vector-like fermions, for instance, $bbb,\; b\tau\tau$, and $\tau\tau\tau$ resonances. Some of the signatures have already been strongly constrained by the existing LHC searches and the parameter space fitting the gamma ray excess is further restricted. On the other hand, the pure $\tau$-rich final state is only weakly constrained so far due to the small electroweak production rate. }

\maketitle

\section{Introduction}

The nature of dark matter (DM) and its interactions are among the biggest puzzles in astro-particle physics. Attempts to understand DM more directly by looking for signals of its interactions with the Standard Model (SM) fields are taking place on a variety of experimental and observational frontiers. Given the current lack of any non-gravitational evidence for DM, we need to consider all theoretically consistent DM possibilities with distinctive experimental signatures. Such an approach can broaden the coverage of search programs by probing new, experimentally unexplored regions of the parameter space. 

One way to classify DM models is by the portals that connect DM with the SM particles. 
Among the simplest possibilities, the portal could be a pseudoscalar. Pseudoscalar portal dark matter (PPDM) has the appealing feature that the interaction it mediates is suppressed by momentum transfer in the $t$-channel process relevant for direct detection experiments, while preserving the weak-scale cross section of the $s$-channel process, which leads to the right amount of thermal relic abundance.

An experimentally inspired motivation for PPDM is that it offers a simple explanation for hints that have been observed in Fermi-LAT gamma ray data. More specifically, data from the Galactic center~\cite{Hooper:2010mq, Hooper:2011ti,Abazajian:2012pn, Daylan:2014rsa, Zhou:2014lva, Calore:2014xka} and the newly discovered dwarf galaxy Reticulum II~\cite{Geringer-Sameth:2015lua, Hooper:2015ula} can be explained by annihilation of DM particles with the mass of a few ten of GeV to about a few hundred of GeV to various SM final states~\cite{Anchordoqui:2013pta, Okada:2013bna, Huang:2013apa, Modak:2013jya, Kyae:2013qna, Boehm:2014hva, Lacroix:2014eea, Hektor:2014kga, Alves:2014yha, Berlin:2014tja, Agrawal:2014una, Abdullah:2014lla, Izaguirre:2014vva, Cerdeno:2014cda, Ipek:2014gua, Ko:2014gha, Ghosh:2014pwa, Han:2014nba, Wang:2014elb, Fields:2014pia, Arina:2014yna, Cheung:2014lqa, Huang:2014cla, Balazs:2014jla, Ko:2014loa, Baek:2014kna, Okada:2014usa, Ghorbani:2014qpa, Bell:2014xta, Banik:2014eda, Borah:2014ska, Cahill-Rowley:2014ora, Yu:2014pra, Guo:2014gra, Cao:2014efa, Agrawal:2014oha, Cheung:2014tha, Calore:2014nla, Kong:2014haa, Biswas:2015sva, Ghorbani:2014gka, Cerdeno:2015ega, Chen:2015nea, Guo:2015lxa, Modak:2015uda, Caron:2015wda, Berlin:2015wwa, Gherghetta:2015ysa, Bi:2015qva, Cline:2015qha, Elor:2015tva, Fortes:2015qka, Ko:2015ioa, Balazs:2015boa, Banik:2015aya, Cao:2015loa, Kim:2015fpa, Jia:2015uea}. The required cross section of DM annihilation is of the same order of magnitude as required to yield the thermal relic abundance matching the observations. 

Gamma ray observations are inherently difficult to interpret in the context of DM, mainly due to astrophysical processes that not only contaminate any faint signal but also can mimic the expected morphological and spectral signatures. For example, even though the gamma ray emission from the Galactic center suggests that the morphology is also  consistent with the expectation from DM~\cite{Daylan:2014rsa}, it is possible that millisecond pulsars are responsible for the observed excess ~\cite{Lee:2015fea, Bartels:2015aea} (for earlier discussions, see~\cite{Hooper:2013nhl, Cholis:2014lta, Calore:2014pca}). However, the situation with dwarf galaxies is different. Dwarf galaxies have always been considered as the cleanest sources for DM annihilation gamma ray searches. The recent gamma ray observations from Reticulum II~\cite{Geringer-Sameth:2015lua, Hooper:2015ula} (but see also \cite{2015arXiv150302632T}) are thus intriguing as they do potentially provide for the first time a hint of a DM signature (something of course that needs to be confirmed with future explorations). 

Annihilation of DM particles in SM channels has many implications for cosmology, the most interesting one being the effects of energy injection prior and during the epoch of recombination \cite{2009PhRvD..80d3526S,2012PhRvD..85d3522F,2013PhRvD..88f3502G,2013PhRvD..87l3513S,2014PhRvD..89j3508M,2015arXiv150603811S}. The effective ionization energy from DM annihilation can  be constrained by CMB measurements, the most recent ones being from the Planck Collaboration \cite{2015arXiv150201589P}. Recent studies \cite{2014PhRvD..89j3508M,2015arXiv150603811S} find that light-lepton final states are excluded for cross sections of order $< {\cal{O}}(10^{-27} {\mathrm{cm}}^3/{\mathrm{s}})$ for annihilations to electron-positron pairs, photons, or $VV \rightarrow 4 e$, where $VV$ is a pair of intermediate vector bosons, each of which decays to an electron-positron pair. Similar constraints on the cross section are obtained for muon final states. 

Given the challenges in  direct, indirect, and cosmological experiments, it will be interesting to explore the power of the Large Hadron Collider (LHC) in testing this simple possibility. Indeed the pseudoscalars could be produced copiously at the LHC and lead to different final states and signals depending on their couplings. If a pseudoscalar ($a$) is heavy enough, its effect could be encoded by dimension-six operators, which have been considered in~\cite{Bai:2010hh, Goodman:2010ku, Goodman:2010yf, Rajaraman:2011wf, Haisch:2012kf, Lin:2013sca} and motivates collider signals with a single ($b$-)jet and missing transverse energy ($\met$). More recently, the particular simplified model in the minimal flavor violating framework with pseudoscalar coupling to the SM fermions proportional to the SM Yukawa couplings has been extensively studied in~\cite{Buckley:2014fba, Harris:2014hga, Haisch:2015ioa, Buchmueller:2015eea}. We will consider two other types of well-motivated simplified models with the pseudoscalar either dominantly coupled to $b$ quarks or $\tau$ leptons among all SM fermions. Since the pseudoscalar couples to both SM particles and DM, it could decay either visibly or invisibly, which calls for different search strategies. One needs to take into account the corresponding branching fractions in evaluating the collider sensitivity to the full parameter space. 

The simple renormalizable pseudoscalar coupling to the SM fermions $i a f_L f_R^c$ is not SM gauge invariant. It originates from dimension-five operators $i a H f_L f_R^c$ with $H$ being the Higgs doublet, $f_L$ the left-handed fermion weak doublet and $f_R$ the right-handed fermion singlet. This means that the pseudoscalar simplified model is generated by integrating out some heavy fields that couple to both SM particles and the pseudoscalar. The coupling between the pseudoscalar and the SM fermions is actually set by the mass scale of the heavy degrees of freedom. The UV completion (at the LHC energy scale) most studied in the literature is the two Higgs doublet model (2HDM) and its variants~\cite{Izaguirre:2014vva, Ipek:2014gua, Berlin:2015wwa}. Yet there exists another simple class of UV completions for the pseudoscalar simplified model, SM augmented by vector-like generations of fermions. We will demonstrate that this class of models provides new decay channels and signatures of vector-like quarks and leptons, such as triple $b$-jet, $b\tau\tau$, and $\tau\tau\tau$ resonances. Some of the signatures have already been probed by current multi-jet or multi-lepton LHC searches. These constraints on the mass scale of the vector-like generations restricts the pseudoscalar coupling to the SM fermions and thus the parameter space that could explain the GeV excess. The pure $\tau$-rich final states are only weakly constrained at the moment. 

The paper is organized as follows: in Sec.~\ref{sec:simplfiedmodels}, we discuss two simplified models of PPDM in which the pseudoscalar couples dominantly either to $b$'s or $\tau$'s and existing experimental bounds on these models. In Sec.~\ref{sec:vector-like}, we construct the class of vector-like fermion models that generate the simplified models and study their new decay channels and LHC signatures. We conclude in Sec.~\ref{sec:conclusion}.

\section{Simplified Models of PPDM}
\label{sec:simplfiedmodels}

The general simplified model of PPDM is 
\beq
{\cal L}_{\rm PPDM}&=& i \left( g_\chi \bar{\chi} \gamma^5 \chi + \sum_f g_f \bar{f} \gamma^5 f \right) a,
\label{eq:simplified}
\eeq
where we assumed DM is a Dirac fermion denoted by $\chi$ and $a$ is the pseudoscalar. The second term sums over all SM fermions (denoted by $f$), except neutrinos. Notice that the operators, $i a \bar{f} \gamma^5 f$ (or equivalently $i a f_L f_R^c + cc.$ in terms of Weyl fermions),  summed in the second term are not SM gauge invariant. They should be understood as dimension-five operators such as $i a H^\dagger \ell_L \tau_R^c/M_*$ and $i a H^\dagger Q_L b_R^c/M_*$ generated from integrating out some new heavy degrees of freedom beyond the SM of mass scale $M_*$. Thus for the simplified model to be valid, $g_f \sim v/(\sqrt{2} M_*)$ with $v=245$ GeV is restricted to be  $\lesssim 1$ (otherwise the higher-dimensional operators start playing an important role and can't be ignored).
We choose to study the simplified models rather than the effective contact interaction operators. The motivation for this choice is because of potential pitfalls of the latter, for instance, the breakdown of the effective operator approach at the LHC for a relatively light mediator with mass of a few hundred GeV~\cite{Busoni:2013lha, Buchmueller:2013dya, Malik:2014ggr}.

\begin{figure*}[t]
\begin{center}
\begin{tikzpicture}[line width=1.5 pt]
\node at (-2.2,1) {$b$};
\node at (-2.2,-1){$g$};
\node at (2.2, 1){$a$};
\node at (2.2, -1){$b$};
\draw[electron] (-2,1)--(0.,1);
\draw[gluon] (-2,-1)--(0,-1);
\draw[electron] (0,-1)--(0,1);
\draw[scalar](0,1)--(2,1);
\draw[electron](0,-1)--(2,-1);
\end{tikzpicture}
\quad
\begin{tikzpicture}[line width=1.5 pt]
\node at (-2.2,1) {$g$};
\node at (-2.2,-1){$g$};
\node at (2.2, 1){$b$};
\node at (2.2, -1){$\bar{b}$};
\node at (2.2, 0) {$a$};
\draw[gluon] (-2,1)--(0.,1);
\draw[gluon] (-2,-1)--(0,-1);
\draw[electron] (0,-1)--(0,1);
\draw[electron](0,1)--(2,1);
\draw[electron](0,-1)--(2,-1);
\draw[scalar](0, 0) -- (2,0);
\end{tikzpicture}
\quad
\begin{tikzpicture}[line width=1.5 pt]
\node at (-2.2,1) {$g$};
\node at (-2.2,-1){$g$};
\node at (2.2, 1){$b$};
\node at (2.2, -1){$\bar{b}$};
\node at (2.2, 0) {$a$};
\draw[gluon] (-2,1)--(0.,1);
\draw[gluon] (-2,-1)--(0,-1);
\draw[electron] (0,-1)--(0,1);
\draw[electron](0,1)--(2,1);
\draw[electron](0,-1)--(2,-1);
\draw[scalar](1, 1) -- (2, 0);
\end{tikzpicture}
 \end{center}
\caption{Representative diagrams for production of a pseudoscalar in the first simplified model.}
\label{fig:pseudoscalardiagrams}
\end{figure*}
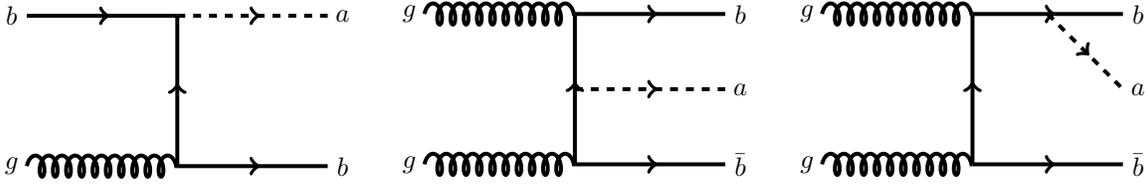

There are many different choices of the couplings between $a$ and the SM fermions. The most studied one in the literature is $g_f  = c \, y_f$, with $c$ a common factor for all SM fermions and $y_f$ the SM Yukawa coupling. In this case, the pseudoscalar dominantly couples to the top quarks and its dominant production channel is the gluon fusion process similar to that of the Higgs boson. The collider phenomenology of this case has been studied in~\cite{Buchmueller:2015eea, Haisch:2015ioa, Harris:2014hga, Buckley:2014fba}. The implications for the GeV excesses has been further studied in~\cite{Buchmueller:2015eea}, which showed that current searches already constrain much of the parameter space that explain the observed excess and that future searches could cover the full parameter space. Here, we  focus on the following two benchmark simplified models:
\begin{enumerate}
\item $g_b \gg g_f\; (f \neq b)$;
\item $g_\tau \gg g_f\; (f \neq \tau)$. 
\label{simplfiedmodels}
\end{enumerate}
The choices represent distinctive collider topologies and search strategies. In the first case, the pseudoscalar is produced associated with $b$-quarks through representative diagrams in Fig.~\ref{fig:pseudoscalardiagrams}. Notice that this set of diagrams is beyond the most studied case with $s$-channel production of the mediator. In the second case, the pseudoscalar is produced through electroweak processes, $ p p \to (Z^*, \gamma^*) \to 2\tau + a$ with a much smaller rate compared to that of a QCD production channel. 
Both of the simplified models could be easily realized in weak-scale models. For example, they all could arise in SM augmented by vector-like generations of fermions with different representations. In addition, the first choice could also be realized in (variants of) Type-II 2HDM with $\tan\beta \gtrsim 10$~\cite{Izaguirre:2014vva, Ipek:2014gua, Berlin:2015wwa} and the second choice could be realized in (variants of) Type-III 2HDM. Below we will discuss the current status of each case. 

In the first simplified model, the pseudoscalar might also be produced through gluon fusion with a heavy-quark loop (e.g., a bottom or top-quark loop). We focus on the associated production assuming that the pseudoscalar-top coupling is negligible, but note that this assumption may not hold in every UV completion model, e.g., 2HDM with a small $\tan\beta < 10$. In that case, the constraints we derive below should be taken as a conservative estimate in those models. The bottom-loop contribution is sub-dominant compared to that of the associated production for values of $g_b$ that could be probed at 8 TeV. 

The pseudoscalar can decay to $\bar{b}b$ and $\bar{\chi}\chi$ with branching fractions determined by $g_\chi$ and $g_b$: 
\beq
\frac{\Gamma(a \to \bar{\chi} \chi)}{\Gamma(a \to \bar{b} b)} = \frac{g_\chi^2}{3g_b^2} \sqrt{1-4\frac{m_\chi^2}{m_a^2}},
\eeq 
where we neglected the $b$-quark mass and the factor of $3$ in the denominator is the color factor. There are two possible collider searches sensitive to this simplified model. One is the CMS search for the pseudoscalar produced in association with $b$-quarks and decaying into a $b$-quark pair~\cite{Khachatryan:2015tra}, which is most constraining when $a$ has a considerable branching fraction to $b$ quarks. 
If $a$ dominantly decays into a pair of DM particles, the most sensitive search could be the one with hard $b$-jets and large missing transverse energy~\cite{Lin:2013sca, Aad:2014vea}. The CMS $Z(bb)H$(inv) boost decision tree analysis could also be sensitive to this case~\cite{Chatrchyan:2014tja}, which we will leave for future explorations. We focus on $m_a >$ 100 GeV. The LHC constraints and future reach for region with $m_a <$ 100 GeV could be found in~\cite{Kozaczuk:2015bea}.

\begin{figure*}[t]
\begin{center}
 \includegraphics[width=0.48\textwidth]{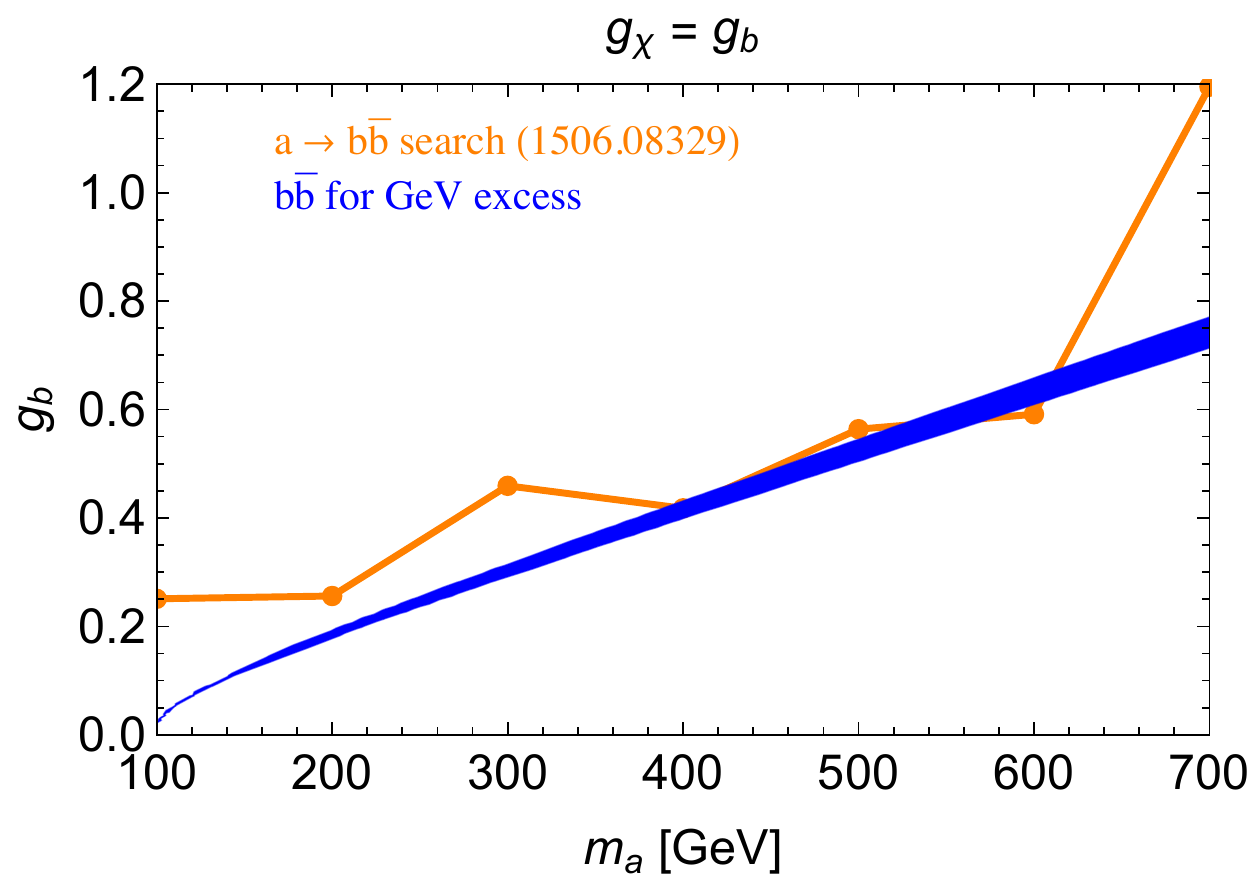} \quad \includegraphics[width=0.48\textwidth]{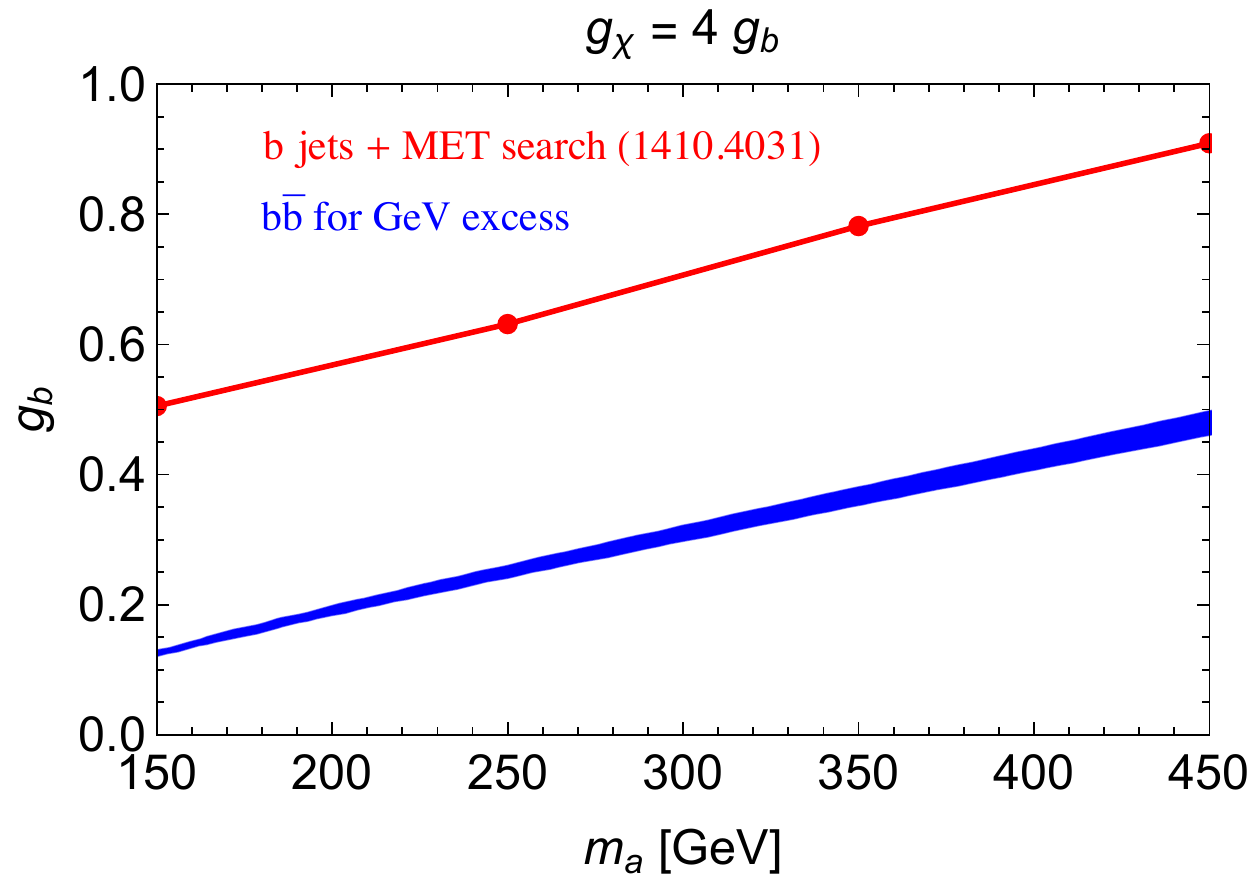}
\end{center}
\caption{A 95\% CL lower bound on the pseudoscalar-$b$ coupling as a function of $m_a$ for $g_\chi = g_b$ (left) and for $g_\chi =4 g_b$ (right). In both panels, the blue contours give the dark matter annihilation cross section that could explain the GeV gamma ray excess in the 1$\sigma$ range. }
\label{fig:monobbound}
\end{figure*}

Figure ~\ref{fig:monobbound} shows the most stringent 95\% confidence level (CL) bound on the DM coupling to $b$-quarks as a function of the pseudoscalar mass from either of the two searches for two representative cases $g_\chi = g_b$ (left) and $g_\chi = 4 g_b$ (right) (at a fixed dark matter mass of 48.7 GeV~\cite{Calore:2014nla}). We also show  the contours with the DM annihilation cross section that could explain the GeV gamma ray excess in the 1$\sigma$ range~\cite{Calore:2014nla}.
We derive the limits in the following way. First we implement the simplified models using FeynRules~\cite{Christensen:2008py}, we then generate parton-level events by feeding the model files into MadGraph5~\cite{Alwall:2011uj}, and finally we use Pythia8~\cite{Sjostrand:2014zea} for parton showering and fragmentation.

We find that for $g_b = g_\chi$, $a$ decays to $b$ quarks with a branching fraction $\gtrsim75\%$, and thus the direct search for $a$ decaying into $b$-quark pairs is more constraining. For $g_\chi = 4 g_b$, we find that as $a$ decays mostly to DM particles with a branching fraction $\sim 85\%$, the DM search with hard $b$-jets is more constraining. We only present result up to $m_a=450$ GeV because at energies beyond 450 GeV current searches are only sensitive to $g_b$ greater than one, where the simplified model description also breaks down.  
We also take into account the pseudoscalar branching fractions when deriving limits (instead of assuming a 100\% branching fraction of $a$ into one particular channel, as compared to previous studies of this simplified model). 
In general, for an arbitrary choice of $(g_b, g_\chi)$, it is important to consider the branching fractions of the pseudoscalar $a$ and study both types of searches in order to derive a robust limit. 
Note that  current searches are only sensitive to about $\gtrsim 3$ times the coupling $g_b$ value required for a DM explanation of the GeV gamma ray excess in most of $a$'s mass range. 

In the second simplified model, the pseudoscalar is produced in association with two $\tau$'s through electroweak processes $ p p \to (Z^*, \gamma^*) \to 2\tau + a \to 4 \tau$ or $2 \tau+\met$. An interesting question is whether the LEP searches were already sensitive to the production of the pseudoscalar through $e^+ e^- \to 2 \tau +a$ process if $a$ is light enough. We find that the production rate of this channel is too small to have been  constrained at LEP. For instance, for $m_a = 50$ GeV and $g_\tau =1$, the production cross section is 13 fb at $\sqrt{s} = 200$ GeV. Given an integrated luminosity of 233.4 pb$^{-1}$ collected per LEP experiment in the year 2000~\cite{Assmann:2002th}, this led to only 3 events per experiment produced in that year. However note that even the current LHC searches are not sensitive to this electroweak process. For $m_a = 50$ GeV and $g_\tau =1$, the production cross section of $a$ is 16.6 fb. If $a$ dominantly decays to 2$\tau$'s, the most sensitive search is the CMS multilepton search~\cite{Chatrchyan:2014aea}, which only excludes 30 times the cross section for $m_a = 50$ GeV. 
Nevertheless,  it is perhaps possible to probe such process through the supersymmetry search with two hadronically decaying $\tau$'s and missing energy if $a$ decays dominantly to DM particles~\cite{Aad:2014yka}. Unfortunately, currently these searches are only sensitive to a production cross section of order a few hundred fb and thus this simplified model is still largely unconstrained. 

If, in the second simplified model, the pseudoscalar has a small but non-negligible coupling to other SM fermions, e.g., bottoms, this will open up new production channel of the pseudoscalar such as the associated production with $b$ quarks. There are constraints from the MSSM Higgs boson search in the $\tau\tau$ final state~\cite{Khachatryan:2014wca, Aad:2014vgg}, which restricts $g_b$ to be below 0.03 for $m_a = 200$ GeV and below 0.1 for $m_a = 700$ GeV assuming 100\% branching fraction of $a \to \tau\tau$.

\section{New Signatures of UV Completions with Vector-Like Fermions}
\label{sec:vector-like}

Concrete models that give rise to the simplified model in Eq.~(\ref{eq:simplified}) could lead to additional interesting experimental signatures. One simple UV completion of the PPDM model, Type-II 2HDM and its variants, have been extensively studied in~\cite{Izaguirre:2014vva, Ipek:2014gua, Berlin:2015wwa}. In this section, we focus on another class of models that realizes PPDM: the vector-like fermion models. We demonstrate that this model could lead to new interesting vector-like fermion signatures that could be worth dedicated searches in the LHC Run 2. 

Consider a generation of vector-like quarks, $B^\prime, \tilde{B}^\prime$ with charge $(3,1)_{-1/3}$ and $(\bar{3},1)_{1/3}$ under SM gauge group $SU(3) \times SU(2)_L \times U(1)_Y$. We  assume that it only couples to the third generation quarks of the SM:
\beq
{\cal L}_{int} = (M_B \tilde{B}^\prime B^\prime + i Y_a a B^\prime b_R^c + y_3 H^\dagger Q_3 \tilde{B}^\prime + cc.) + i a \bar{\chi} \gamma^5 \chi, \label{eq:vectorlike}
\eeq
where $H$ is the SM higgs doublet and $Q_3$ is the third generation SM quark doublet. Integrating out the $B^\prime$ field leads to a dimension-five operator 
\beq
i\frac{Y_a y_3}{M_B} H^\dagger Q_3  b_R^c a + cc. ,
\eeq
which gives a $(Y_a y_3v / \sqrt{2} M_B)a b_L b_R^c$ coupling with $v = 245$ GeV after electroweak symmetry breaking (EWSB). 

\begin{figure}[h]
\begin{center}
\begin{tikzpicture}[line width=1.5 pt]
\node at (-1.2,1) {$p$};
\node at (-1.2,-1) {$p$};
\node at (2.0,0.8) {$B^\prime$};
\node at (2.0,-0.8) {$B^\prime$};
\node at (3.5, 1.2) {$a$};
\node at (3.5, -1.2) {$a$};
\node at (3.5,2.2){$b$};
\node at (5.0,2.2){$\bar{b}$};
\node at (5.7,1){$b$};
\node at (3.5,-2.2){$\bar{b}$};
\node at (5.0,-2.2){$b$};
\node at (5.7,-1){$\bar{b}$};
\draw[electron] (-1,1)--(0.,0);
\draw[electron] (-1.,-1)--(0,0);
\draw[gluon] (0,0)--(1.5,0);
\draw[electron] (1.5,0)--(2.5,1);
\draw[electron] (1.5,0)--(2.5,-1);
\draw[electron] (2.5,1)--(3.5,2);
\draw[scalar] (2.5,1)--(4.0,1);
\draw[electron] (4.0,1)--(5.5,1);
\draw[electron] (4.0,1)--(5.0,2);
\draw[electron] (2.5,-1)--(3.5,-2);
\draw[scalar] (2.5,-1)--(4.0,-1);
\draw[electron] (4.0,-1)--(5.5,-1);
\draw[electron] (4.0,-1)--(5.0,-2);
\end{tikzpicture}
\quad
 \includegraphics[width=0.45\textwidth]{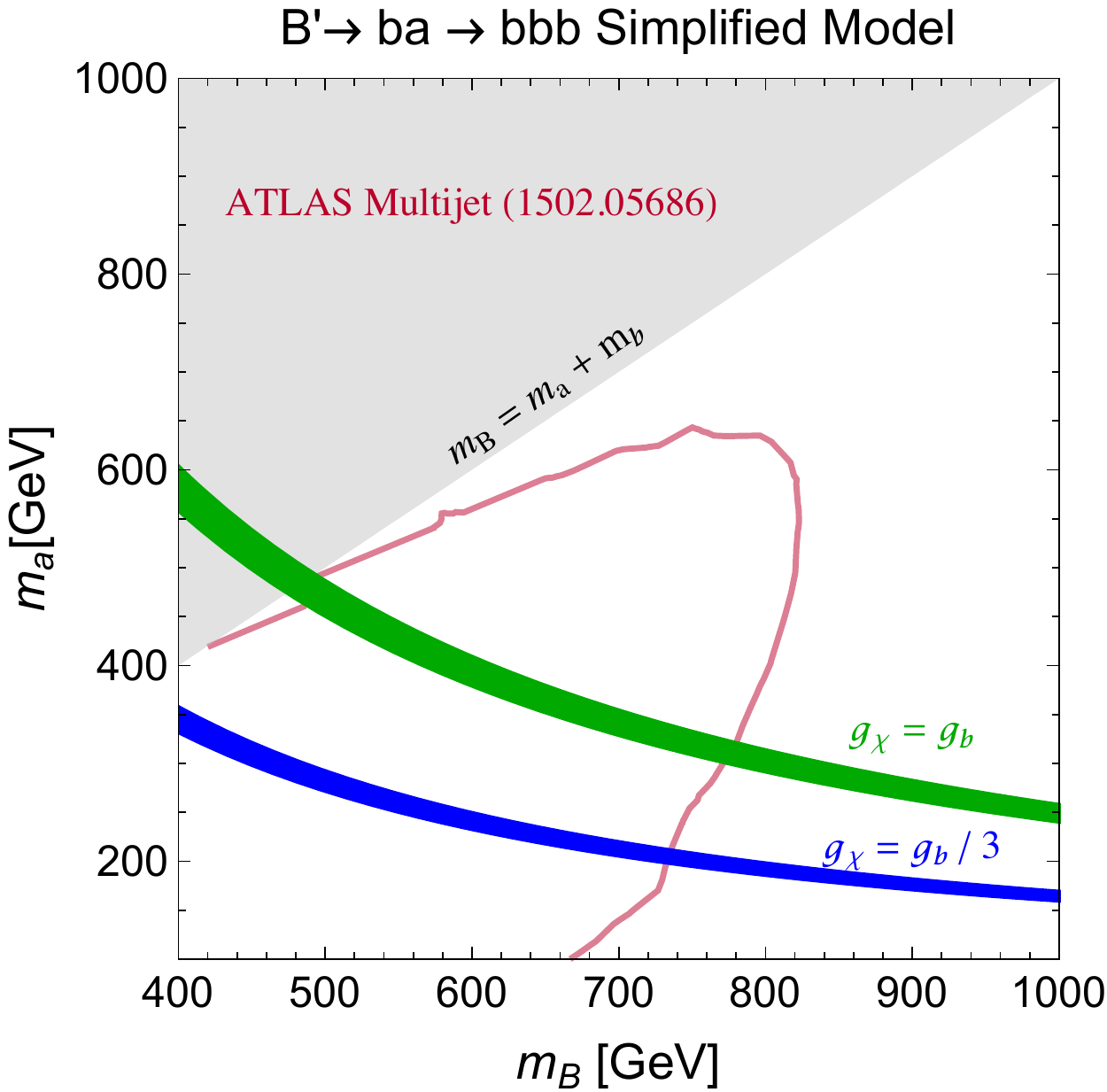}
 \end{center}
\caption{Left: diagram of $B^\prime$ pair production followed by $B^\prime \to a b, a \to b \bar{b}$ decay. Right: bound on the vector-like $B^\prime$ quarks decaying through $B^\prime \to a b, a \to b \bar{b}$ with 100\% branching fraction based on the results of the jet counting analysis in the multijet search in~\cite{Aad:2015lea} (red solid curve). The green and blue bands give the cross section of DM annihilation fitting the GeV gamma ray excess in the $b\bar{b}$ channel within the 1$\sigma$ range. We fixed $g_\chi$ to be $g_b$ (green) or $g_b/3$ (blue) so that $a \to b \bar{b}$ is the dominant decay channel and $g_b = v/M_B$, which is equivalent to setting $Y_a y_3 = \sqrt{2}$.}
\label{fig:3jetbound}
\end{figure}
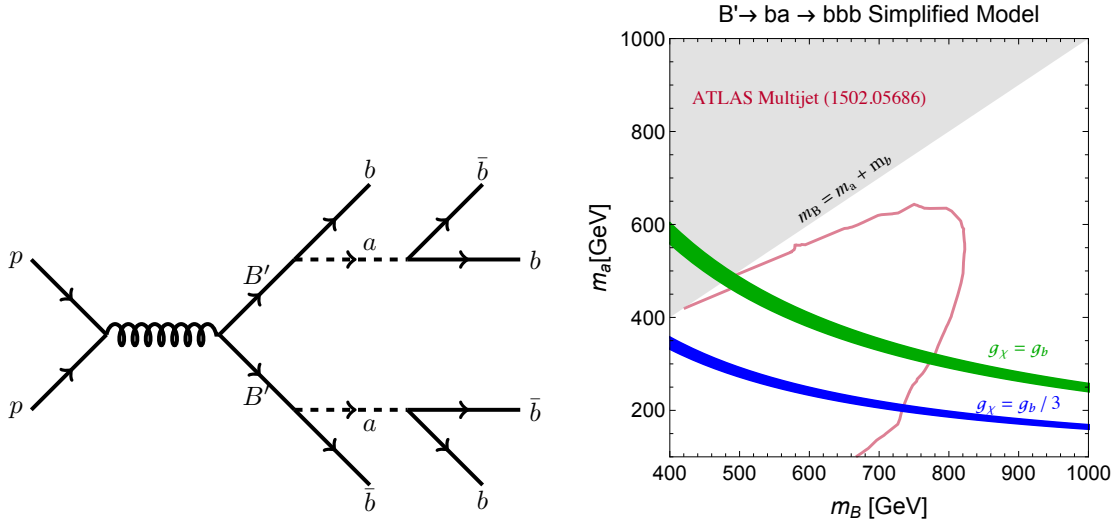

Note that the second operator in Eq.~(\ref{eq:vectorlike}) opens up a new decay channel of the new quarks, $B^\prime \to a b$ with a partial width 
\beq
\Gamma(B^\prime\to a b) = \frac{Y_a^2 m_B}{48 \pi} \left(1- \frac{m_a^2}{m_B^2}\right)^2.
\eeq
The partial widths of the other standard decay channels $B^\prime \to Wt, Zb, hb$ are proportional to $y_3^2$ with a ratio $\Gamma(B^\prime \to Wt): \Gamma(B^\prime \to Zb): \Gamma(B^\prime \to hb) = 2:1:1$, up to some small phase space corrections. 
For $Y_a > y_3$, $B^\prime \to a b$ could become the dominant decay channel of the new fermions\footnote{A $y_3$ of order one contributes negatively to the running of the Higgs quartic coupling and could lead to a potential vacuum instability problem which appears at around 10 - 100 TeV~\cite{Blum:2015rpa} while either $y_3$ or $Y_a$ of order one have Landau poles at above a PeV. The model should be taken as a valid effective field theory at energy scales LHC could probe.}\footnote{In addition, Yukawa couplings between the Higgs boson and the new fermions modify various Higgs couplings and electroweak precision observables. These constraints are sensitive to the quantum numbers of the new fermions. For instance, adding weak singlet vector-like new fermions are strongly constrained by the $Zb\bar{b}$ coupling measurements yet it could be embedded in custodial models such as Refs.~\cite{Agashe:2006at, Batell:2012ca}. We want to separate the collider studies from the indirect probes and focus on the new collider signatures. We refer the readers to~\cite{Blum:2015rpa} and references therein for a detailed discussion of indirect constraints on vector-like fermions.}. This leads to an interesting new signature of the vector-like $B^\prime$ quarks: 6 $b$-jets from pair production of $B^\prime$'s with the correct combination of jet triplets with the invariant masses near the $B^\prime$ mass and one of the three possible jet pairs within each triplet with the invariant mass close to that of the pseudoscalar (as shown in Fig.~\ref{fig:3jetbound}, left). In general, the signature of a $b$-jet-rich final state (with more than two $b$-jets and small amount of missing transverse energy) could arise in a variety of new physics scenarios. In addition to the vector-like fermion model we discuss here, it could also arise in several supersymmetric scenarios, e.g. $R$-parity violating and ``stealth" supersymmetry~\cite{Fan:2011yu, Evans:2014gfa, Fan:2012jf}. Consequently, it could be worthwhile to perform a dedicated multi-$b$-jet search to probe all these different scenarios. 

There are two current searches sensitive to this topology. One is the jet counting analysis in the ATLAS multijet search~\cite{Aad:2015lea}. The analysis requires a large number of hard jets combined with a requirement on the number of $b$-tagged jets. In particular, the signal region with $n_{\rm jet} \geq 7$, $p_T^{\rm jet} \geq 80$ GeV and $n_{\rm b-tags}=2$ is most sensitive to the signal we are interested in in the context of this work. The 95\% CL limit on the number of events that new physics contributes to this region is 38. Another sensitive search is the CMS search for $R$-parity violating gluino decaying to three jets~\cite{Chatrchyan:2013gia}. This search uses jet ensemble technique~\cite{Essig} to look for pair-produced three-jet resonances in multijet events from $pp$ collisions. The analysis looks for both light- and heavy-flavor jets. The heavy-flavor jet search region requires that each event contains at least one $b$-tagged jet, which increases the signal significance and thus sets stronger constraint.
 
To recast the experimental results, we implement the model described by Eq.~\ref{eq:vectorlike} using FeynRules, then feed the model files into MadGraph5 to generate the parton-level events, and finally use Pythia8~\cite{Sjostrand:2014zea} to describe parton shower and fragmentation. In this  analysis, we use anti-$k_T$ clustering algorithm incorporated in FastJet~\cite{Cacciari:2011ma} to define jets. The cross section of $B^\prime$ pair production was taken from~\cite{vectorlikexsec, Aad:2015kqa}, which is an NNLO result. We validate our analysis codes using the sample of $R$-parity violating gluino decays. The constraints on the vector-like quark from the ATLAS jet counting analysis is shown in the right panel of Fig.~\ref{fig:3jetbound}.  Assuming 100\% branching fraction of $a \to b\bar{b}$, the current constraint excludes the new fermion in the mass range 700--820 GeV, which is comparable to the results of searches assuming $B^\prime$ decaying to $Wt$, $Zb$ or $hb$~\cite{vectorlike}. Note that the CMS jet resonance search requires a fitting to the shape of the jet triplet invariant mass distribution, which we approximated with a mass window selection, yielding comparable, yet slightly weaker bounds on the $B^\prime$ mass.

To understand the implications of the constraints for a potential DM explanation of the GeV gamma ray excess from the Galactic center and Reticulum II, we also plot on right panel of Fig.~\ref{fig:3jetbound} bands yielding the cross section of DM annihilating into $b\bar{b}$ that could explain the excess~\cite{Calore:2014nla}. We fixed $g_\chi$ to be $g_b$ or $g_b/3$ so that $a \to b \bar{b}$ is the dominant decay channel and $g_b = v/M_B$, which is equivalent to setting $Y_a y_3 = \sqrt{2}$. One could see that the constraints on $B^\prime$ forces the pseudoscalar to be lighter than 300 GeV in order to explain the GeV gamma ray excess. Then the light $a$ could be probed by a direct mediator search as discussed in Sec. 2.

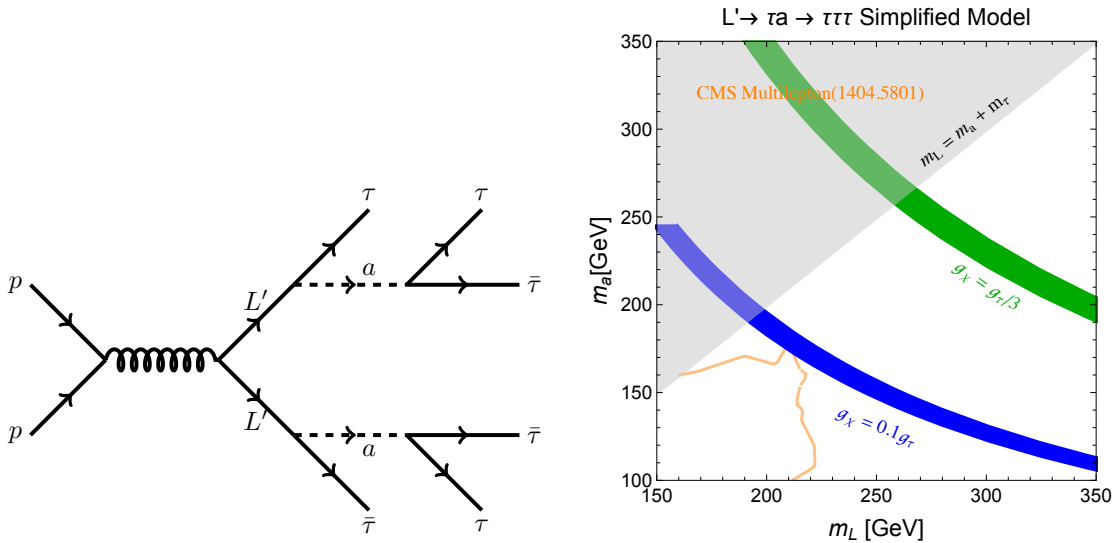
\begin{figure}[h]
\begin{center}
\begin{tikzpicture}[line width=1.5 pt]
\node at (-1.2,1) {$p$};
\node at (-1.2,-1) {$p$};
\node at (2.0,0.8) {$L^\prime$};
\node at (2.0,-0.8) {$L^\prime$};
\node at (3.5, 1.2) {$a$};
\node at (3.5, -1.2) {$a$};
\node at (3.5,2.2){$\tau$};
\node at (5.0,2.2){$\tau$};
\node at (5.7,1){$\bar{\tau}$};
\node at (3.5,-2.2){$\bar{\tau}$};
\node at (5.0,-2.2){$\tau$};
\node at (5.7,-1){$\bar{\tau}$};
\draw[electron] (-1,1)--(0.,0);
\draw[electron] (-1.,-1)--(0,0);
\draw[gluon] (0,0)--(1.5,0);
\draw[electron] (1.5,0)--(2.5,1);
\draw[electron] (1.5,0)--(2.5,-1);
\draw[electron] (2.5,1)--(3.5,2);
\draw[scalar] (2.5,1)--(4.0,1);
\draw[electron] (4.0,1)--(5.5,1);
\draw[electron] (4.0,1)--(5.0,2);
\draw[electron] (2.5,-1)--(3.5,-2);
\draw[scalar] (2.5,-1)--(4.0,-1);
\draw[electron] (4.0,-1)--(5.5,-1);
\draw[electron] (4.0,-1)--(5.0,-2);
\end{tikzpicture}
\quad
 \includegraphics[width=0.45\textwidth]{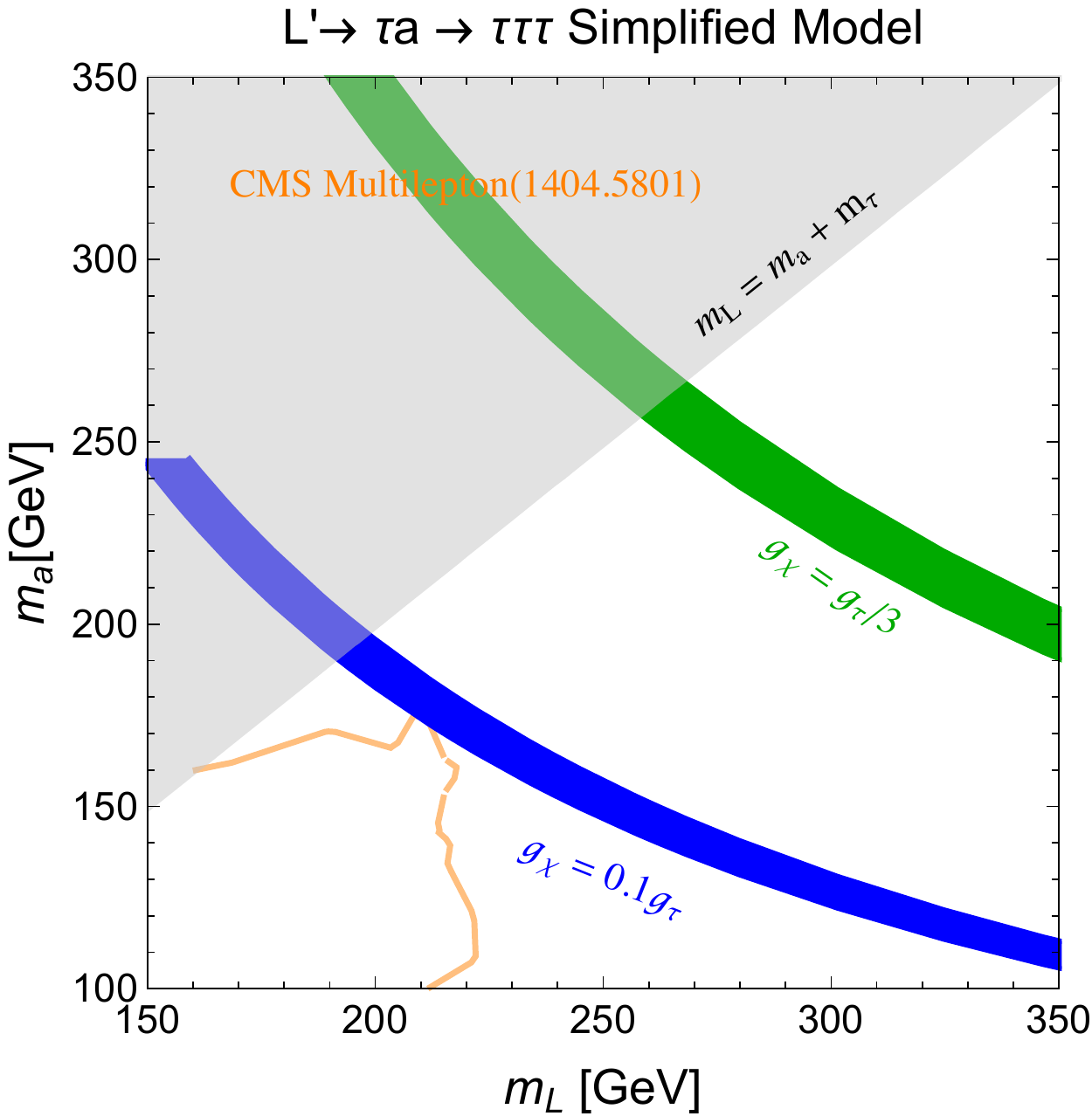}
 \end{center}
\caption{Left: diagram of $L^\prime$ pair production followed by $L^\prime \to a \tau, a \to \tau \bar{\tau}$ decay. Right: bound on the vector-like $L^\prime$ quarks decaying through $L^\prime \to a \tau, a \to \tau \bar{\tau}$ with 100\% branching ratio based on the results of the multilepton search in~\cite{Chatrchyan:2014aea} (orange solid curve). The green and blue bands give the cross section of DM annihilation fitting the GeV excess in the $\tau\bar{\tau}$ channel within the 1$\sigma$ range. We fixed $g_\chi$ to be $g_\tau/3$ (green) or $0.1g_\tau$ (blue) so that $a \to \tau \bar{\tau}$ is the dominant decay channel and $g_\tau = v/M_L$, which is equivalent to setting $Y_a y_3 = \sqrt{2}$.}
\label{fig:3taubound}
\end{figure}

One could also consider the SM augmented with a vector-like generation of leptons. One simplest possibility is $L^\prime, \tilde{L}^\prime$ with charge $(1,1)_{\pm1}$ under the SM gauge group $SU(3) \times SU(2)_L \times U(1)_Y$. We  assume that it only couples to the $\tau$ leptons in the SM with interactions
\beq
{\cal L}_{int} = (M_L \tilde{L}^\prime L^\prime + i Y_a a L^\prime \tau_R^c + y_3 H^\dagger \ell_3 \tilde{L}^\prime +cc.) + i a \bar{\chi} \gamma^5\chi, \label{eq:vectorlike-L}
\eeq
with $\ell$ being the lepton doublet in the SM. Similar to the vector-like $B^\prime$ model, integrating out the heavy leptons leads to an effective coupling between $a$ and $\tau$ leptons. This also leads to a new decay channel of $L^\prime$: $L^\prime \to \tau a \to \tau \tau \bar{\tau}$. Considering $L^\prime$ pair production, this leads to a 6$\tau$ final state with the two correct $\tau$ triplet combinations each forming a resonance in the invariant mass spectrum. In addition, one of the three $\tau$ pair combinations in each of the two triplets should also form an invariant mass peak at the $a$ mass. Currently, the most sensitive experimental probe of this scenario is the CMS multilepton search~\cite{Chatrchyan:2014aea}. More specifically, the search regions with one hadronically decaying $\tau$ and a total number of leptons $\geq 4$ are the most sensitive ones to the topology of interest. The results are shown in Fig.~\ref{fig:3taubound}. Assuming 100\% branching fraction of $a \to \tau^+{\tau^-}$, the current constraint only excludes the new leptons below 230 GeV and it is only sensitive to a part of the parameter space that could explain the GeV gamma ray excess. 

\begin{figure}[h]\begin{center}
\begin{tikzpicture}[line width=1.5 pt]
\node at (-1.2,1) {$p$};
\node at (-1.2,-1) {$p$};
\node at (2.0,0.8) {$B^\prime$};
\node at (2.0,-0.8) {$B^\prime$};
\node at (3.5, 1.2) {$a$};
\node at (3.5, -1.2) {$a$};
\node at (3.5,2.2){$b$};
\node at (5.0,2.2){$\tau$};
\node at (5.7,1){$\bar{\tau}$};
\node at (3.5,-2.2){$b$};
\node at (5.0,-2.2){$\tau$};
\node at (5.7,-1){$\bar{\tau}$};
\draw[electron] (-1,1)--(0.,0);
\draw[electron] (-1.,-1)--(0,0);
\draw[gluon] (0,0)--(1.5,0);
\draw[electron] (1.5,0)--(2.5,1);
\draw[electron] (1.5,0)--(2.5,-1);
\draw[electron] (2.5,1)--(3.5,2);
\draw[scalar] (2.5,1)--(4.0,1);
\draw[electron] (4.0,1)--(5.5,1);
\draw[electron] (4.0,1)--(5.0,2);
\draw[electron] (2.5,-1)--(3.5,-2);
\draw[scalar] (2.5,-1)--(4.0,-1);
\draw[electron] (4.0,-1)--(5.5,-1);
\draw[electron] (4.0,-1)--(5.0,-2);
\end{tikzpicture}
\quad
 \includegraphics[width=0.45\textwidth]{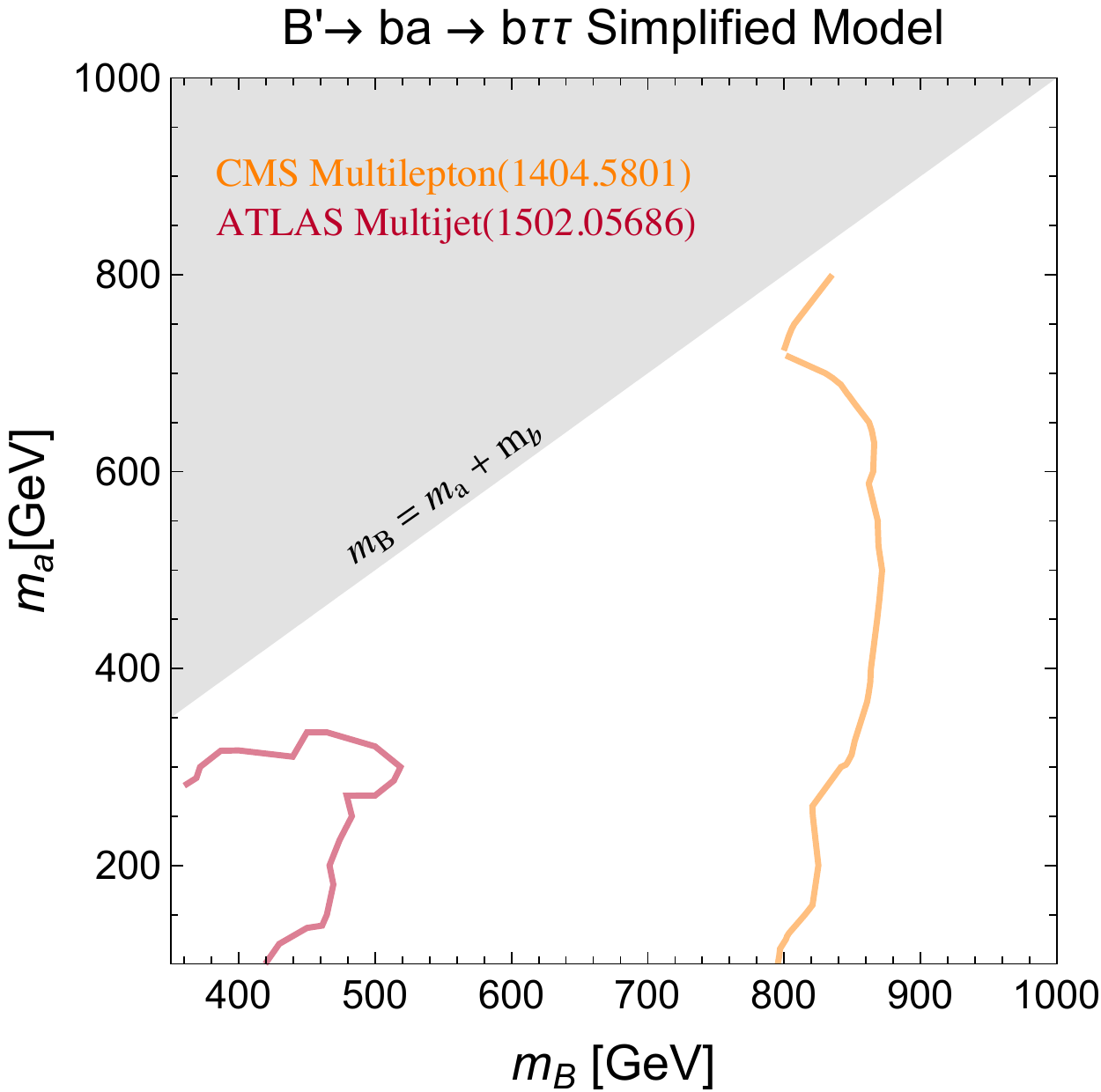}
 \end{center}
\caption{Left: diagram of $B^\prime$ pair production followed by $B^\prime \to a b, a \to \tau \bar{\tau}$ decay. Right: bound on the vector-like $B^\prime$ quarks with the decay topology $B^\prime \to a b, a \to \tau \bar{\tau}$ based on the multijet search in~\cite{Aad:2015lea} (red) and multilepton search in~\cite{Chatrchyan:2014aea} (orange).}
\label{fig:taubound}
\end{figure}
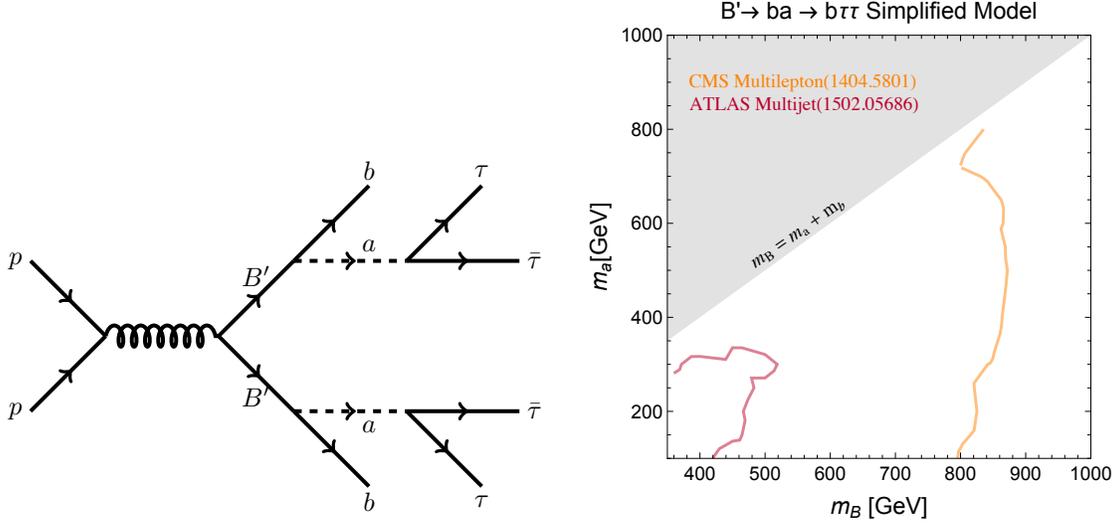

Combining the two simplified models above, one could have SM augmented with vector-like $B^\prime$ quarks, as well as vector-like leptons $L^\prime$, both of which couple to the third-generation particles. Pseudoscalar $a$ couples to both $B^\prime$ and $L^\prime$. This could lead to a new possibility of $B^\prime$ decay: $B^\prime \to a b \to b \tau \bar{\tau}$ as shown in the left diagram in Fig.~\ref{fig:taubound}. In this case, assuming 100\% branching fraction, we find a strong constraint on this decay topology from the CMS multilepton search~\cite{Chatrchyan:2014aea}, which excludes $B^\prime$ below 800 GeV. This is shown in the right panel of Fig.~\ref{fig:taubound}. We also check the multijet search constraint on this topology and find a much weaker limit, as demonstrated in the same figure. 

\begin{figure}[h]\begin{center}
\begin{tikzpicture}[line width=1.5 pt]
\node at (-1.2,1) {$p$};
\node at (-1.2,-1) {$p$};
\node at (2.0,0.8) {$T^\prime$};
\node at (2.0,-0.8) {$T^\prime$};
\node at (3.5, 1.2) {$a$};
\node at (3.5, -1.2) {$a$};
\node at (3.5,2.2){$t$};
\node at (5.0,2.2){$\tau$};
\node at (5.7,1){$\tau$};
\node at (3.5,-2.2){$\bar{t}$};
\node at (5.0,-2.2){$\tau$};
\node at (5.7,-1){$\tau$};
\draw[electron] (-1,1)--(0.,0);
\draw[electron] (-1.,-1)--(0,0);
\draw[gluon] (0,0)--(1.5,0);
\draw[electron] (1.5,0)--(2.5,1);
\draw[electron] (1.5,0)--(2.5,-1);
\draw[electron] (2.5,1)--(3.5,2);
\draw[scalar] (2.5,1)--(4.0,1);
\draw[electron] (4.0,1)--(5.5,1);
\draw[electron] (4.0,1)--(5.0,2);
\draw[electron] (2.5,-1)--(3.5,-2);
\draw[scalar] (2.5,-1)--(4.0,-1);
\draw[electron] (4.0,-1)--(5.5,-1);
\draw[electron] (4.0,-1)--(5.0,-2);
\end{tikzpicture}
\quad 
 \includegraphics[width=0.45\textwidth]{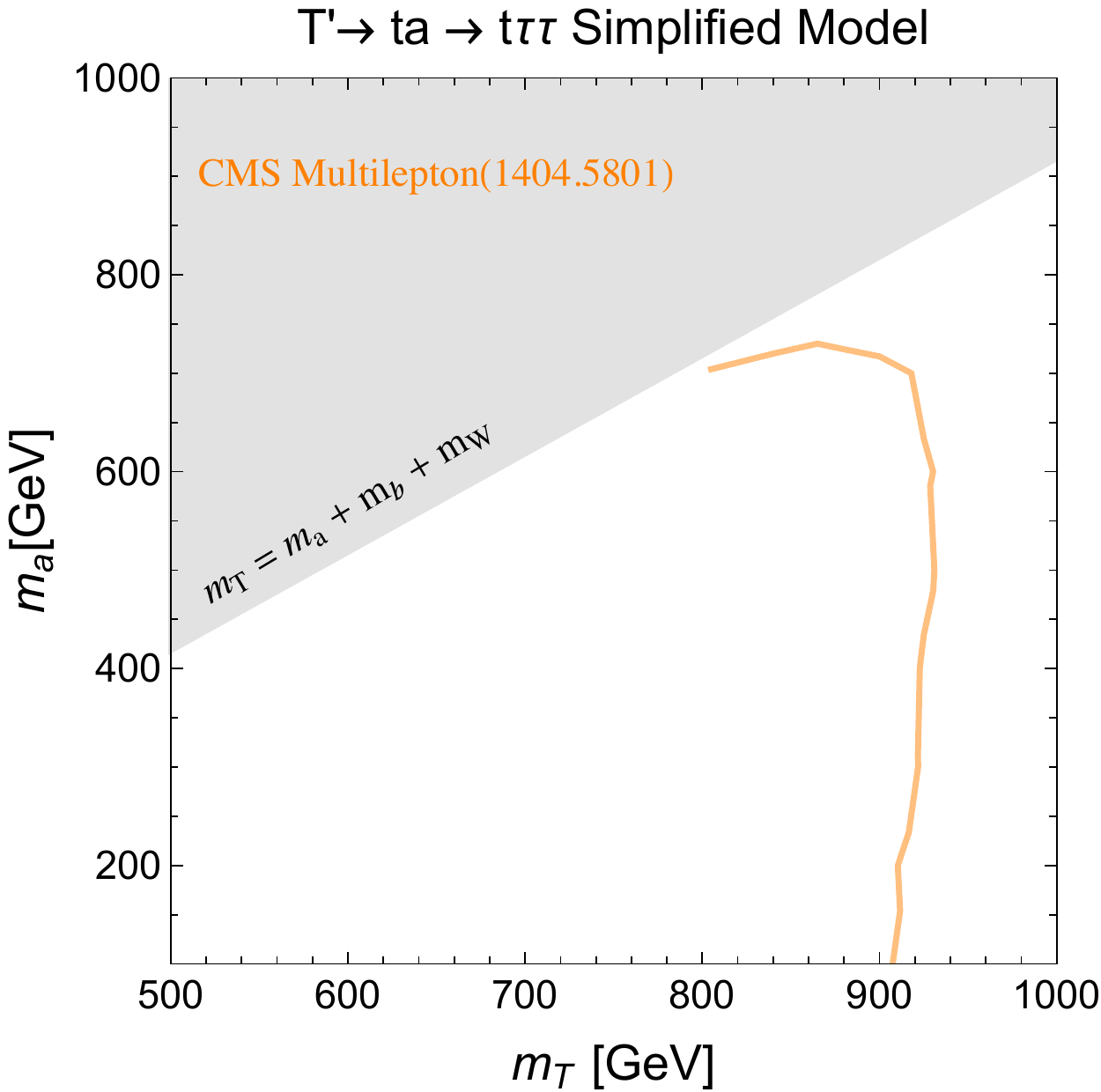}
\end{center}
\caption{Left: diagram of $T^\prime$ pair production followed by $T^\prime \to a t, a \to \tau \bar{\tau}$ decay. Right: bound on the vector-like $T^\prime$ quarks with the decay topology $T^\prime \to a t, a \to \tau \bar{\tau}$ based on the multilepton search in~\cite{Chatrchyan:2014aea} (orange). }
\label{fig:ttaubound}
\end{figure}
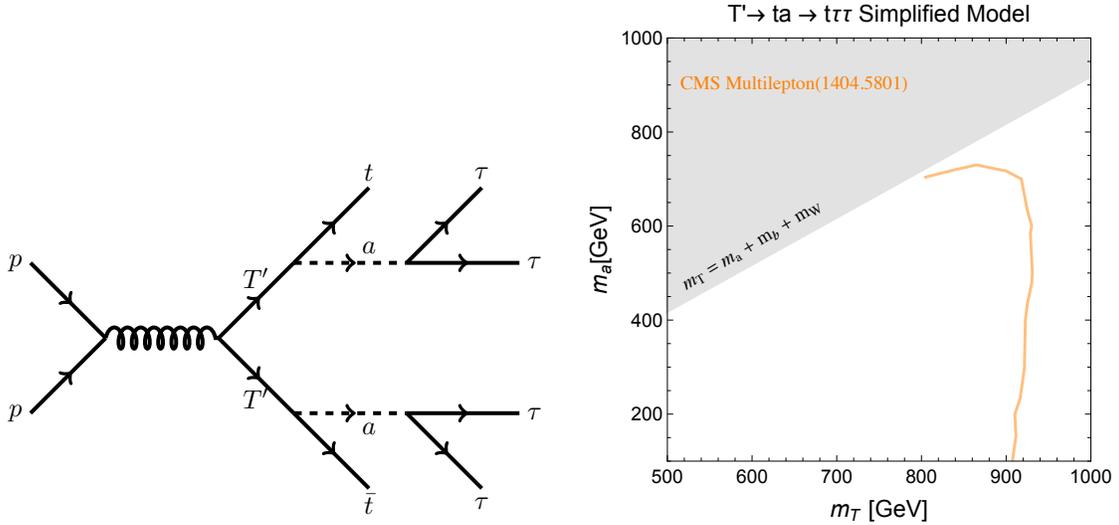

Depending on the representations of the new fermions, there could be more new signatures of vector-like fermions beyond the SM. 
For instance, in a model with a pair of new quarks with charge $(3,2)_{1/6}$ and $(\bar{3},\bar{2})_{-1/6}$, the new top-quark partner ($T^\prime$) could decay to $a+t $ with a final state of $ t \bar{t} + 4 b$ jets. This decay channel of $T^\prime$ is studied in~\cite{Anandakrishnan:2015yfa} that shows that current searches still allow for a lighter top quark partner with the mass as low as 500 GeV for a certain range of $m_a$. Again in slightly more complicated models, it is possible to have $T^\prime$ decaying to $a+t$ followed by $a \to \tau^+{\tau^-}$, which leads to a $t \bar{t} + 4\tau$ final state. The strongest constraint in this case comes from the CMS multilepton search~\cite{Chatrchyan:2014aea}, which excludes $T^\prime$ with the mass up to about 900 GeV, as shown in Fig.~\ref{fig:ttaubound}. Note that both Fig.~\ref{fig:taubound} and Fig.~\ref{fig:ttaubound} could still be compatible with the GeV gamma ray excess, depending on the $L^\prime$'s parameters.

It is also of interest to embed a simplified phenomenological model of vector-like fermions in a more elaborate EWSB model, such as little Higgs models~\cite{ArkaniHamed:2002qy, ArkaniHamed:2002qx}. 
For example, it has been shown that littlest Higgs model could contain a pseudoscalar with the top-quark partner decaying dominantly to it and the top-quark in at least part of the parameter space~\cite{Kilian:2004pp}. More recently, more models containing the pseudoscalar and new decay channels of top quark partner in the broad context of composite EWSB have been proposed in~\cite{Anandakrishnan:2015yfa, Serra:2015xfa}. 

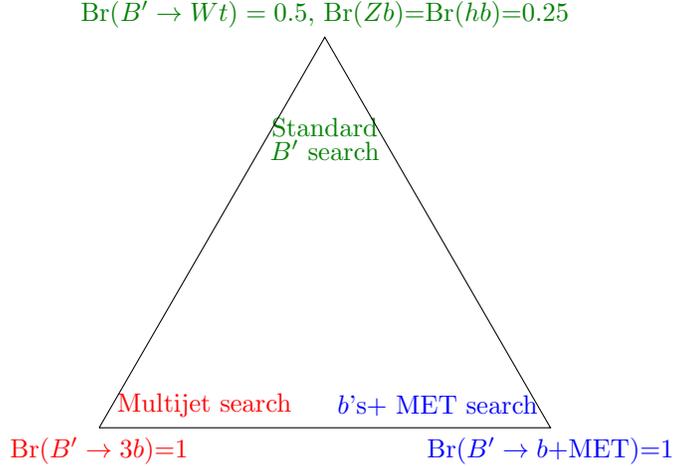
\begin{figure}[h]\begin{center}
\begin{tikzpicture}
\node [red] at (0,-0.3) {Br$(B^\prime \to 3b)$=1};
\node [red] at (1.4, 0.3) {Multijet search}; 
\node [black!50!green] at (3,5.5) {Br$(B^\prime \to Wt) = 0.5$, Br$(Zb)$=Br$(hb)$=0.25};
\node [black!50!green] at (3, 4.0) {Standard };
\node [black!50!green] at (3, 3.7) {$B^\prime$ search};
\node [blue] at (6, -0.3) {Br$(B^\prime \to b$+MET)=1};
\node [blue] at (4.5, 0.3) { $b$'s+ MET search};
\draw (0,0) -- (3,5.1962) -- (6,0)-- (0,0);
\end{tikzpicture}
\end{center}
\caption{New triangle of $B^\prime$ search for model described by Eq.~\ref{eq:vectorlike}. }
\label{fig:new triangle}
\end{figure}

Finally, we want to comment that all the collider constraints demonstrated in this section are based on the assumption that a particular decay chain has a 100\% branching fraction. In principle, multiple decay topologies exist for a given model. For instance, in the $B^\prime$ model, since the pseudoscalar also couples to DM particles, $B^\prime$ could decay through $B^\prime \to b a \to b + \met$. We could then have the asymmetric final states with $B^\prime \to bbb$ in one decay chain and $B^\prime \to b + \met$ in the other. It is rather useful to present these search results in a triangle with Br$(B^\prime \to Wt, Zb, hb)$, Br$(B^\prime \to 3b)$, and Br$(B^\prime \to b+\met$) =1 at the three vertices. At each vertex, the most sensitive search is quite different, as we demonstrate in Fig.~\ref{fig:new triangle}.

\section{Conclusions}
\label{sec:conclusion}
In this paper, we explore two simplified models of pseudoscalar portal dark matter and a class of its UV completion with vector-like fermions. We showed that for pseudoscalar dominantly coupled to $b$-quarks, the most sensitive searches could be either the direct search for scalar particles decaying into $b$-quark pairs or the $b$-jets+$\met$ searches. The current sensitivity still allows for a dark matter explanation of the GeV gamma ray excesses in the Galactic center and/or Reticulum II. If the pseudoscalar dominantly couples to $\tau$'s, current LHC searches are not sensitive to it due to the small electroweak production rate. The pseudoscalar coupling to SM could be realized in SM augmented by vector-like fermions with the coupling strength set by the mass scale of the new fermions. These new vector-like models give rise to interesting signatures such as $bbb, b\tau\tau, \tau\tau\tau$ resonances. Considering pair production of the new quarks, the 6 $b$-jet and 2$b$+4$\tau$ final states have already been strongly constrained by the current multijet or multilepton searches. Thus in the vector-like quark model, the parameter space for dark matter explanation of GeV excess is further constrained. On the other hand, pair production of new leptons with 6$\tau$ final state (with two triple-$\tau$ resonances, and also resonances in one of the di-$\tau$ pairs in each triplet) is only weakly constrained due to the small production cross section. 
It will be interesting to explore these types of simplified models and new vector-like fermion signatures further in the upcoming LHC runs.

\section{Acknowledgements}
\label{sec:acknowledgements}
We acknowledge useful conversations with Gordon Blackadder, Alex Geringer-Sameth, Matt Reece and Matthew Walker.
The work of GL and SMK is partially supported by the DOE Award DE-SC0010010-003376; SMK is also supported by NSF PHYS-1417505 and NASA NNX13AO94G Awards. JF would like to acknowledge the hospitality of the Aspen Center for Physics(supported by NSF PHYS-1066293), where part of this work has been conducted.  
 
\bibliography{ref1.bib}
\bibliographystyle{utphys}

\end{document}